\documentclass[twoside,twocolumn]{article}

\usepackage{blindtext} 

\usepackage[sc]{mathpazo} 
\usepackage[T1]{fontenc} 
\usepackage{microtype} 

\usepackage[english]{babel} 

\usepackage[top=3cm,bottom=3cm,left=2cm,right=2cm]{geometry}

\usepackage{booktabs} 

\usepackage{enumitem} 
\setlist[itemize]{noitemsep} 

\usepackage{abstract} 

\usepackage{titlesec} 
\titleformat{\section}[block]{\large\scshape\centering}{\thesection.}{1em}{} 
\titleformat{\subsection}[block]{\large}{\thesubsection.}{1em}{} 

\usepackage{fancyhdr} 
\usepackage{titling} 

\usepackage{hyperref} 
\usepackage{graphicx}

\pretitle{\begin{center}\Huge\bfseries} 
\posttitle{\end{center}} 
\title{Observation of the full spiral spectrum of a light beam with single-pixel detection} 

\author{%
\textsc{Luis Jos\'{e} Salazar-Serrano$^{1}$, Job Mendoza-Hern\'{a}ndez$^{1}$, Juan P. Torres$^{1,2}$} \\[1ex] 
\normalsize $^{1}$ICFO - The Institute of Photonic Sciences, the Barcelona Institute of Science and Technology\\ \normalsize 08860 Castelldefels (Barcelona), Spain \\ \normalsize $^{2}$Dep. Signal Theory and Communications, Universitat Politecnica de Catalunya\\ \normalsize Campus Nord D3, 08034 Barcelona, Spain \\
\normalsize Corresponding author: juanp.torres@icfo.eu 
}
\date{\today} 

\renewcommand{\maketitlehookd}{%
\begin{abstract}
\normalfont{We demonstrate experimentally a {\em simple} and {\em easy-to-use} technique aimed at measuring the complex orbital angular momentum spectrum of an arbitrary optical field making use of just polarization measurements. The technique can be applied to any other family of modes used to describe the spatial shape of optical beams. The idea is inspired by a formal analogy that exist between classical fields that can show certain amount of correlations between different degrees of freedom (polarization and spatial shape) and the degree of entanglement that might exist between two subsystems, so it is an example of a {\em quantum-inspired} classical technology.}
\end{abstract}
}

\begin{document}
\renewcommand{\abstractname}{\vspace{-\baselineskip}}

\maketitle


\section{Introduction}
Tailoring of the the spatial shape of light beams is an important
resource that is being considered nowadays for many novel photonic
applications \cite{book2011,roadmap}. In general, the spatial
shape cannot be considered separately from the polarization (spin)
\cite{jackson1999,cohen1999}. However, in the paraxial regime both
contributions can be measured and manipulated independently, so
the description of the spatial shape of a light beam can be
considered separately from its polarization.

In this scenario, the spatial distribution of the intensity and
phase of the optical field, $F(x,y)$, even down to the single
photon limit, can be written as a mode decomposition
\begin{equation}
F(x,y)=\sum_m C_m \,G_{m} (x,y)\,,
\end{equation}
where $C_m=\int dx\, dy \, F(x,y) G_{m}^*(x,y)$, $G_{m}(x,y)$ is a
set of functions that constitute a basis and the array $\left\{C_m
\right\}$ is the set of complex numbers that uniquely describe
$F(x,y)$.

In many applications \cite{communications,witness,wai_pan2015} one
chooses as basis the set of Laguerre-Gauss modes,
$\varphi_{l,p}(x,y)$, where the modes are labeled by two indexes
$m=l,p$ and each mode carries an orbital angular momentum (OAM) of
$m\hbar$ per photon \cite{allen1992}. In this case, the array
$\left\{C_{lp} \right\}$ is the so-called spiral spectrum of the
light beam \cite{torres2003}. Notwithstanding, a family of modes
might exist that is more convenient to describe the optical field
transformations that take place, by and large.

In principle, we can think of obtaining full information about the
spatial shape of the beam, amplitude and phase of the optical
field at each point, by measuring these quantities directly with a
variety of methods available that require the use of a CCD camera.
Afterwards one can calculate any mode decomposition of the optical
field computationally. For instance, phase-shifting digital
holography \cite{phase_shiting} measures the spatial shape of the
optical field by measuring the intensity resulting from the
interference of the field with a reference field with four
different phase offsets.

However, these types of procedures based on the use of {\em
multiple pixels} (cameras) might not be very convenient in most
practical applications. This is especially true at wavelengths
where there are no good cameras or these are complicated, bulky
and expensive \cite{single_pixel}. In this scenario it might be
preferable to measure directly the weights of the corresponding
mode decomposition, or the weights of specific modes of interest,
which can be done with the help of a single-pixel camera. This
would make easier the measurement of optical fields at wavelengths
where mega-pixel cameras are not affordable.

There are also applications where the information of interest
about an object can be retrieved observing specific modes
contained in the optical field. This is the case, for instance of
Digital Spiral Imaging \cite{DSI,willner}, where one aims at
probing certain properties of a sample by inspecting how certain
spatial modes are transformed by its interaction with the sample.
In this case it seems indeed unnecessary to obtain full
information of the field.

Several techniques aimed at elucidating the orbital angular
momentum spectrum of arbitrary optical fields has been put forward
and demonstrated over the the years. Vasnetsov et al.
\cite{vasnetsov2003} showed that the OAM spectrum of light beams
can be retrieved measuring the different frequency shifts induced
by the rotation of each OAM mode contained in the field around the
beam axis \cite{courtial1998,zhou2017}. However rotating a beam
about its own axis at a fast rate is generally difficult.

One can concatenate interferometers that sort out different OAM
states, although due to its interferometric nature this technique
is technically demanding \cite{leach2002}. Another approach is to
make use of a technique that transforms helically phase beams to
tilted plane waves, where the tilt depends on the OAM index, and
then measure the intensity at different positions with the help of
a CCD camera \cite{berkhout2010}. With a similar technique
Trichili et al. \cite{carmelo2016} were able to identify with high
fidelity 105 encoded modes where each mode manifest as separate
spatial locations after transmission through a properly engineered
transmission filter.

By projecting the incoming optical field into a specific mode
$\varphi_{l,p}(x,y)$, with the help of holograms or computer-controlled
spatial light modulators (SLM), and detecting the outgoing
intensity, one can measure the OAM spectrum $\left\{ |C_m|^2
\right\}$. Notice that this method provides no information about
the phase of the spectrum, which may be relevant in certain
applications. The measurement of the phase necessitates projecting
into combinations of modes \cite{forbes2013}.

One drawback of these projection-based measurements is that one
needs to perform a sequence of projective measurements at
different times, hence the time to characterize an optical field
increases with the dimension of its modal spectrum, which makes
difficult the measurement of fields with large dimensions.
Recently a technique aimed at overcoming this limitation and that
performs direct measurements of complex fields with extremely high
dimensionality have been demonstrated \cite{boyd1,boyd2}.

Here we put forward and demonstrate a {\em simple} scheme to
measure the full complex spectrum of an arbitrary optical field.
The idea (see Fig. 1) is to generate, with the help of a
polarizing beam splitter, a combined state of polarization and
spatial shape of the form $\Psi(x,y)=F(x,y) {\bf H}+ G_{m} (x,y)
{\bf V}$, and perform only polarization measurements of the
outgoing signal. $F(x,y)$ is the optical field to be
characterized, $G_m(x,y)$ is the basis and ${\bf H}$ and ${\bf V}$
designate horizontal and vertical polarizations, respectively.

The technique described here is a clear example of how quantum
insights can inspire novel classical technologies. The idea of our
technique comes from noticing the similarity of the classical
field
\begin{equation}
\Psi(x,y)=F(x,y) {\bf H}+ G (x,y) {\bf V}\,,
\end{equation}
with the quantum state of a single photon
\cite{adam,nagali2009,karimi2010}
\begin{equation}
| \Psi \rangle=\frac{1}{\sqrt{2}} \left[ a_{f,{\bf H}}^{\dag}+
a_{g,{\bf V}}^{\dag} \right] |vac \rangle\,,
\end{equation}
where $a_{f,{\bf H}}^{\dag}$ is the creation operator of a photon
with spatial shape $f(x,y)$ and horizontal polarization; similarly
$a_{f,{\bf V}}^{\dag}$ is the creation operator for a photon with
spatial shape $g(x,y)$ and vertical polarization.

If one would apply the quantum concepts of purity and concurrence
\cite{wootters1998} to the states given by Eqs. (2) or (3), the
purity would read as
\begin{equation}
P=(1+|\epsilon|^2)/2\,,
\end{equation}
and the concurrence
\begin{equation}
C=|\epsilon|\,,
\end{equation}
where
\begin{equation}
\epsilon=\int \,dx \,dy \,F(x,y) G^{*} (x,y)\,,
\end{equation}
is the overlap between the fields $F$ and $G$. In this scenario of
formal similarities, the entropy of entanglement would read as
\begin{equation}
S=-\lambda_1 \log_2 \lambda_1- \lambda_2 \log_2 \lambda_2\,,
\end{equation}
where
\begin{equation}
\lambda_{1,2}=\frac{1}{2} \left( 1 \pm  |\epsilon| \right)\,.
\end{equation}
If $F$ and $G$ are orthogonal there is maximum {\em entanglement}
of 1 ebit, and the polarization state is an incoherent
combinations of vertical and horizontal polarization. If they are
almost equal, the entanglement is close to zero and the state of
polarization is $1/\sqrt{2} \left( {\bf H} + {\bf V}\right)$. We
make use these {\em quantum-classical} formal analogies to devise
a technique to obtain the complex OAM spectrum of an arbitrary
optical field with just polarization measurements.

\begin{figure}[t]
\centering
\includegraphics[width=0.6\linewidth]{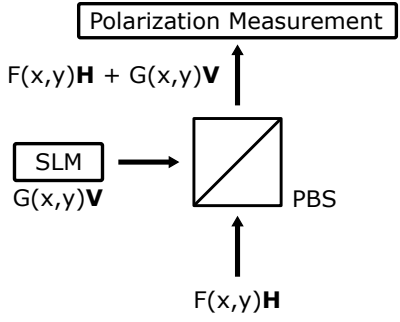}
\caption{Sketch of the technique aimed at measuring the complex
overlap between two light beams with spatial profiles $F(x,y)$ and
$G(x,y)$. The input beam with spatial profile $F(x,y)$ and
horizontal polarization is combined with a second beam with
vertical polarization and spatial profile $G(x,y)$. The state of
polarization of the outgoing signal is measured, which gives the
sought-after overlap between both functions.} \label{fig:1}
\end{figure}

\section{Theoretical background}
The scheme, described by Fig.\ref{fig:1}, is implemented
experimentally by embedding the optical field of interest $F(x,y)$
with horizontal polarization and generating with an SLM a family
of modes $G_{m}(x,y)$ that constitute a basis to describe an
arbitrary optical field. The modes of the basis show vertical
polarization. The outgoing signal $\Psi(x,y)$ after the polarizing
beam splitter (PBS) is given by Eq. (2), and can be written as
\begin{equation}
\Psi(x,y)=\frac{{\bf D}+{\bf A}}{\sqrt{2}} F(x,y)+ \frac{{\bf
D}-{\bf A}}{\sqrt{2}} G_m(x,y)\,,
\end{equation}
where the diagonal polarization reads as ${\bf D}=({\bf H}+{\bf
V})/\sqrt{2}$ and the anti-diagonal polarization reads as ${\bf
A}=({\bf H}-{\bf V})/\sqrt{2}$. The power detected when the
outgoing signal is projected into the ${\bf D}$ and ${\bf A}$
polarization states are
\begin{eqnarray}
& & P_D=\frac{1}{2} \left[ I_F+I_{G_m}+2 {\cal R} \left( C_m
\right)\right]\,,
\nonumber \\
& & P_A=\frac{1}{2} \left[ I_F+I_{G_m}-2 {\cal R} \left( C_m
\right) \right] \nonumber\,,
\end{eqnarray}
where ${\cal R}$ designates the real part. Therefore,
\begin{eqnarray}
{\cal R} \left( C_m \right)=\frac{P_D-P_A}{2}\,. \label{eq:10}
\end{eqnarray}
Similarly, if we project the signal into circularly-polarized
polarization states ${\bf R}=({\bf H}+i{\bf V})/\sqrt{2}$ and
${\bf L}=({\bf H}-i{\bf V})/\sqrt{2}$,
\begin{eqnarray}
& & P_R=\frac{1}{2} \left[ I_F+I_{G_m}-2 {\cal I} \left( C_m
\right)\right]\,,
\nonumber \\
& & P_L=\frac{1}{2} \left[ I_F+I_{G_m}+2 {\cal I} \left( C_m
\right) \right]\,, \label{eq:11}
\end{eqnarray}
where ${\cal I}$ designates the imaginary part. Finally we obtain
\begin{eqnarray}
{\cal I} \left( C_m \right)=\frac{P_L-P_R}{2}\,. \label{eq:12}
\end{eqnarray}
Notice that we measure ${\cal R} \left( C_m \right)$ and ${\cal I}
\left( C_m \right)$ by performing polarization measurements only.
One can thus take advantage of the enormous set of high-quality
optical components existing aimed at manipulating and measuring
optical polarization.

\section{Experimental results}
\subsection{Experimental scheme}
\begin{figure}[t]
\centering
\includegraphics[width=\linewidth]{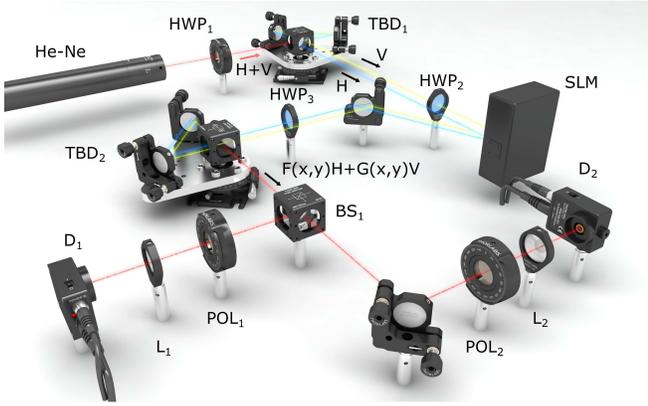}
\caption{Experimental scheme. An input beam with Gaussian profile
is split spatially into two beams with orthogonal polarizations
($\mathrm{TBD_1}$). The beams with horizontal and vertical
polarizations acquire different spatial beam profile given by
$F(x,y)$ and $G_m(x,y)$, respectively, by means of a SLM. After
combining ($\mathrm{TBD_2}$) the beams and performing a
polarization measurement, the overlap between $F(x,y)$ and
$G_m(x,y)$ is measured. HWP: Half-Wave plate; TBD: Tunable Beam
Displacer; SLM: Spatial Light Modulator; BS: Beam Splitter
(50:50); POL: Polarizer; L: Lens; D: detector.} \label{fig:2}
\end{figure}

The experimental scheme is shown in figure\,\ref{fig:2}. A He-Ne
laser (Melles Griot, $\lambda=632.8\mathrm{nm}$) generates a
Gaussian beam polarized at $+45^{\circ}$ with the help of a Half
Wave Plate, $\mathrm{HWP}_{1}$. Afterwards, the beam is split into
two beams that propagate parallel and that are spatially separated
a distance $d$, by means of a Tunable Beam Displacer \cite{TBD},
$\mathrm{TBD}_{1}$. The TBD is a device composed of a Polarizing
Beam Splitter (PBS) and an L-shaped platform with two fixed
mirrors arranged in a Sagnac like configuration. The platform is
free to rotate with respect to the PBS center with the help of a
standard rotating stage. When the platform is rotated at a certain
angle $\theta$, an input beam polarized at $+45^{\circ}$ is
divided into two parallel beams whose separation can be
continuously tuned (i.e. $d \propto \theta$). The two output beams
are linearly polarized with either horizontal or vertical
polarization and no optical path difference is introduced between
them. The wavelength dependence of the TBD, the maximum separation
between the output beams and its spatial quality are mainly
determined by the PBS characteristics.

The two output beams leaving $\mathrm{TBD}_{1}$ impinge on
different areas of the Spatial Light Modulator (SLM). The screen
of the SLM is divided into two separate sections that imprint a
different spatially varying phase after reflection to each
incoming beam. In Fig.~\ref{fig:2}, the rightmost section
introduces a phase profile so that the output beam spatial beam
profile is described by the function $F(x,y)$. Similarly, the left
section generates a beam described by the function $G_m(x,y)$.
Notice that since the SLM can only operate on input beams with
horizontal polarization, a second Half Wave Plate,
$\mathrm{HWP}_{2}$, is introduced before the SLM in the path of
one of the incoming beams in order to rotate its polarization from
vertical to horizontal and impinge on the SLM with the required
polarization.

After the SLM, $\mathrm{TBD}_{2}$ combines the two spatially
separated beams into a single beam. In order to guarantee that the
second TBD operates in inverse mode (with respect to
$\mathrm{TBD}_{1}$) the polarization state of one of the input
beams is rotated by $90^{\circ}$, by a third HWP
($\mathrm{HWP}_{3}$) so that two input beams separated by a
distance $d$ and with orthogonal polarizations are the input.
After $\mathrm{TBD}_{2}$, the output beam is composed of two
collinear beams with orthogonal polarizations. At this point, the
optical field reads
\begin{equation}
\Psi(x,y) = F(x,y)\,{\bf H} + G_m(x,y)\exp(i\phi){\bf V}\,,
\label{eq:1}
\end{equation}
where $\phi$ is a phase introduced by the SLM aimed at i)
compensating any unwanted phase due to misalignment and ii) it
allows to measure the imaginary part of $C_{m}$ by replacing
$\phi$ for $\phi+\pi/2$.

In order to measure the real part of the overlap $C_{m}$, the
optical signal is divided in two components with equal amplitudes.
One component is projected into a diagonal (${\bf D}$)
polarization state and the other into an anti-diagonal (${\bf A}$)
polarization state. After projection, the power measurements of
detectors $\mathrm{D_{1}}$ and $\mathrm{D_{2}}$ are subtracted. In
the experiment, measurements are performed by two detectors
Thorlabs PDA36A-EC connected to a NI-USB 6009 data acquisition
card. The whole measurement procedure is controlled using a Python
script that generates for each SLM window a set of phase masks
that describe the functions $F(x,y)$ and $G_m(x,y)$, varies the
phase $\phi$, performs the power measurements and records the data
obtained.

Following a similar procedure, the imaginary component of $C_m$ is
measured after adding a phase of $\pi/2$ to the current value of
$\phi$. This is achieved by increasing the minimum gray level on
the left window of the SLM. In order to measure the full spiral
spectrum of the beam described by the function $F(x,y)$, two sets
of measurements are performed. In the first set, the real part is
determined and is recorded as a function of the basis index $m$
corresponding to $G_{m}(x,y)$. In the second set, the imaginary
part is measured by performing the same scan over the index $m$
after adding $\pi/2$ to $\phi$. The whole measurements provides
information about the complex character of $\{C_{m}\}$.

\subsection{Results}
\begin{figure}[t]
\centering
\includegraphics[width=\linewidth]{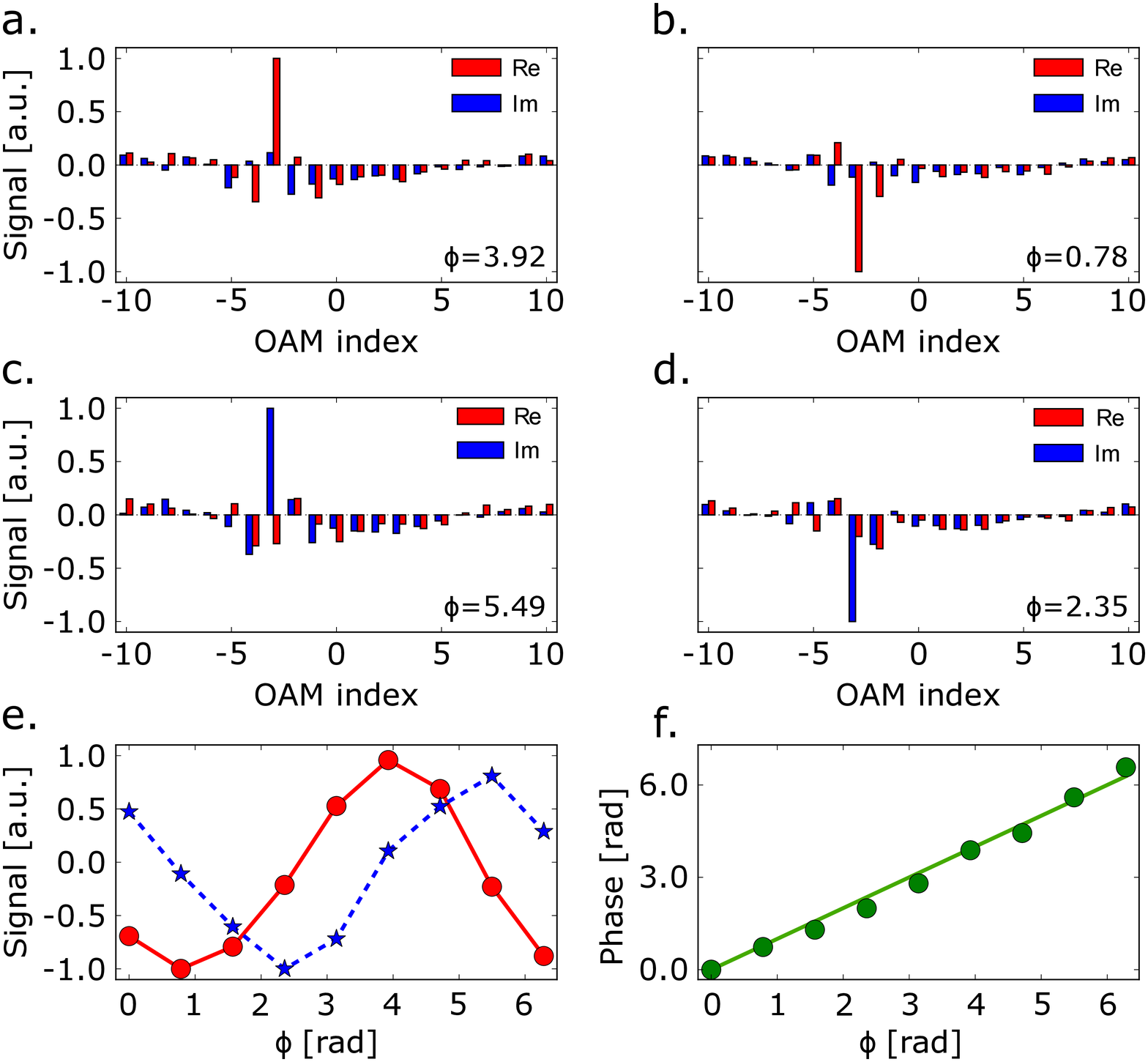}
\caption{Spiral spectrum of a LG beam  with OAM index l=-3 (see
text for details). Notice the change of sign of the signal in
panels (a) - (b), and panels (c) - (d), due to the different
choice of the phase $\phi$. In (e) we show the real (dots) and
imaginary (stars) parts of the overlap function $C_{l=-3}$ as a
function of $\phi$ for $[0,2\pi]$. (f) shows the phase of $C_l$
obtained from the curves in (e) (dots: experiment; line: theory).}
\label{fig:3}
\end{figure}

In order to validate the technique described here, we consider
different optical fields $F(x,y)$ and measure its overlap ($C_l$)
with a set of LG modes with $p=0$. As a first example,
Fig.~\ref{fig:3} considers the decomposition of a LG beam with OAM
index $l=-3$ and varying phase $\phi$, i.e.,
$F(x,y)=\varphi_{l=-3}(x,y)\exp(i\phi)$, in terms of 21 LG modes
with $l\in[-10,10]$. Figs 3(a) to 3(d) depict the real and
imaginary parts of $C_l$ for (a) $\phi=3.92$, (b) $\phi=0.78$, (c)
$\phi=5.49$ and (d) $\phi=2.35$. The data recorded shows a peak
centered in $l = -3$, as expected and the amplitude of the real
and imaginary components, indicated by the red and blue bars
respectively, varies according to the phase $\phi$ as expected.

However the main peak is accompanied by other minor peaks at $l
\ne -3$ with small amplitudes. These secondary peaks appear due to
the fact that the spatial modes at the output of
$\mathrm{TBD_{2}}$ are not pure LG modes. The PBS present in each
TBD degrades significantly the spatial quality of any input beam.
In particular, we have noticed that the spatial quality of the
reflected LG beams varies significantly with respect to the
transmitted counterpart.

In addition to this, the amplitude of these secondary peaks is
also related to the quality of the $50:50$ beam splitter
$\mathrm{BS_{1}}$ positioned after $\mathrm{TBD_{2}}$. Since the
reflected and transmitted components are not split equally, the
first and second terms of $P_R$ and $P_L$ in Eq.\,(\ref{eq:11})
does not fully cancel after subtraction. As an example, for the BS
used in the experiment (Thorlabs pellicle BS CM1-BP150), the
reflected component with horizontal polarization has an amplitude
reduction of $10\%$ to $20\%$ with respect to the other beams,
namely, reflected beam with vertical polarization and transmitted
beam with horizontal and vertical polarizations.

\begin{figure}[t]
\centering
\includegraphics[width=\linewidth]{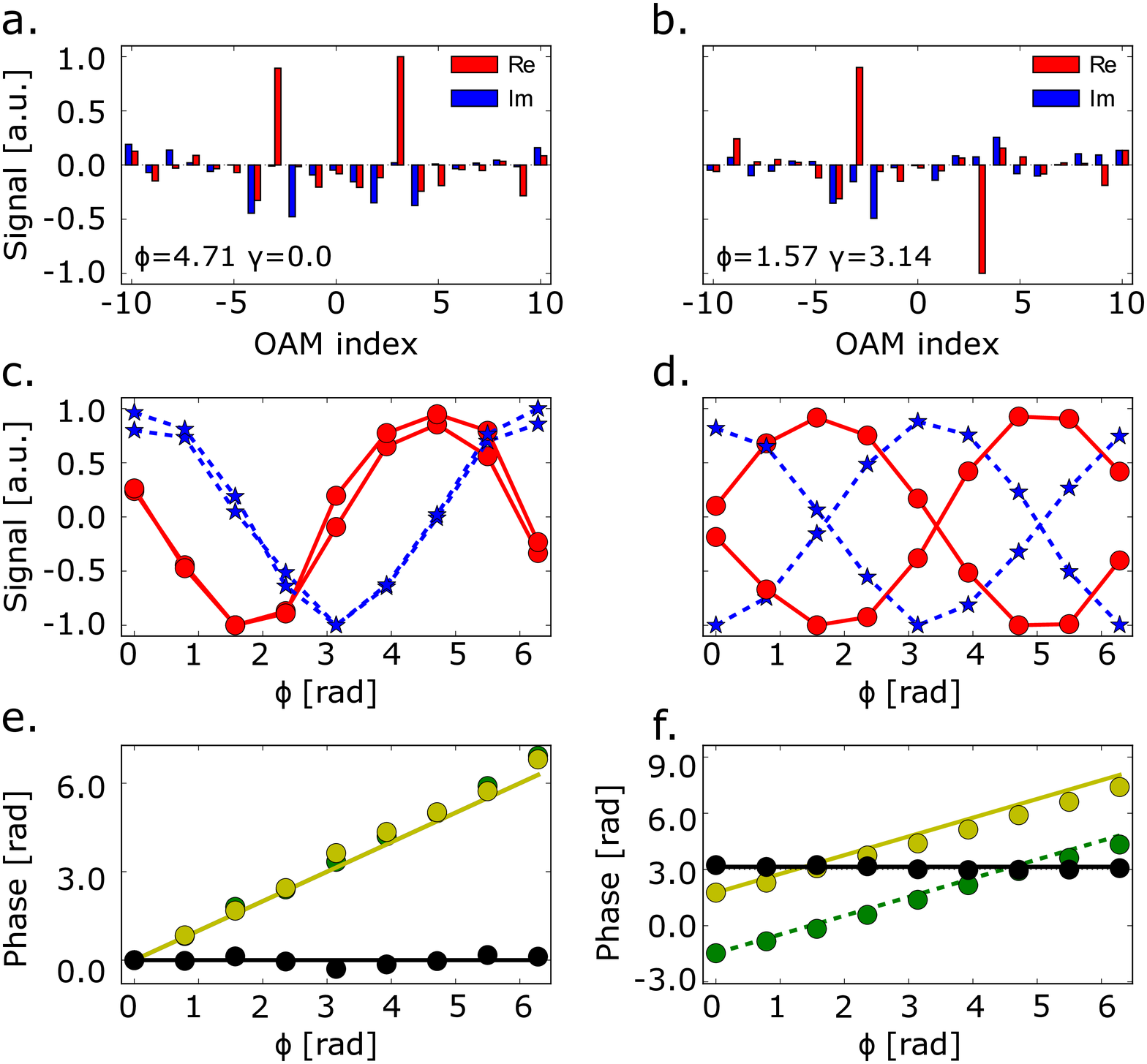}
\caption{Spiral spectrum of a superposition of two LG beams with
OAM indexes $l=+3$ and $l=-3$. (a), (c) and (e): the two LG beams
have the same phase ($\gamma=0$), and (b), (d) and (f): the two LG
beams have a relative phase of $\gamma=\pi$. (c) and (d) show the
real (dots) and imaginary (stars) parts of the overlap functions
$C_{l=-3}$ and $C_{l=+3}$ as a function of $\phi$ for $[0,2\pi]$.
(e) depicts the phase as a function of $\phi$ and the phase
difference between mode projections. Notice that, even though
$\phi$ increases, the difference remains zero as expected (black
line). Similarly, (f) shows the phase of each overlap function
when $\gamma=\pi$. In this case, the black line indicates a
difference in phase of $\sim 3.14$, as expected (dots:experiment;
lines: theory).} \label{fig:4}
\end{figure}

Fig.~\ref{fig:3}(e) shows the real (circles) and imaginary (stars)
parts of the signal measured as a function of the varying phase
$\phi$. Notice that even though $F(x,y)$ and $G_l(x,y)$ are real
functions, the measured real component is maximum for a non-zero
phase due to the fact that an additional phase is introduced due
to misalignments on the TBDs. Finally Fig.~\ref{fig:3}(f) shows
the phase calculated from the real and imaginary values measured
in Fig.~\ref{fig:3}(e), as a function of the phase $\phi$. The
continuous line indicates the expected phase measured as a
function of the phase introduced by the SLM experimentally.

Fig.~\ref{fig:4} presents experimental results for a more general
case where the optical fields is $F(x,y) \sim \left[
\varphi_{l=+3}(x,y)+ \exp(i\gamma)\varphi_{l=-3}(x,y) \right]
\exp(i\phi)$, where $\gamma$ corresponds to a relative phase
between the two modes. Figs.~\ref{fig:4}(a) and (b) present the
measured overlap $C_l$ for two selected cases. In both cases, two
main peaks centered at $l=-3$ and $=+3$ are observed. For
$\phi=4.71$ (Fig.~\ref{fig:4}(a)), the real part of $C_{l=+3}$ is
maximum. The peaks corresponding to $l=-3$ and $l=+3$ have the
same sign and very similar amplitude. Fig.~\ref{fig:4}(c) shows
the overlaps $C_{l=-3}$ and $C_{l=+3}$ measured (real and
imaginary parts) as a function of the phase $\phi$. The phases of
$C_{l=-3}$ and $C_{l=+3}$ are plotted in Fig.~\ref{fig:4}(e). As
expected, since $\gamma=0$, although the phases of  $C_{l=-3}$ and
$C_{l=+3}$ varies as $\sim \exp(i\phi)$, the relative phase
between both overlaps functions is zero (horizontal line).

To illustrate the capability of the technique presented,
Fig.~\ref{fig:4}(b), (d) and (f) present the corresponding results
for $\gamma=\pi$. Fig.~\ref{fig:4}(b) shows that $C_{l=-3}$ and
$C_{l=+3}$ have opposite signs.  Fig.~\ref{fig:4}(f) shows the
measured phases of $C_{l=-3}$ and $C_{l=+3}$ as well as its
difference. The plot confirms that the phase difference between
modes $\varphi_{l=+3}$ and $\varphi_{l=-3}$ (horizontal line) is
around $3.1\,\mathrm{rad}$, which corresponds to $\pi$.

\section{Conclusions}
We have demonstrated a technique that allows to measure the
complex mode spectrum of an arbitrary optical field with just
polarization measurements. The technique, {\em simple} and {\em
easy-to-use}, retrieves the real and imaginary parts of the
overlap of the optical field under investigation with all the
modes of the basis.

In order to obtain the full mode spectrum, the technique should
scan sequentially all the modes of the basis, which can be time
consuming when analyzing optical fields with very high
dimensionality. However, new technological advances should appear
\cite{ferroelectric} that might increase substantially the
refreshing time of spatial light modulators, reducing therefore
the time required to obtain the spectrum.

Finally, we notice that this technique is an example of how
quantum insights can inspire novel classical technologies. The
idea presented and demonstrated comes from noticing the formal
similarity existing between classical fields that might show
correlations between degrees of freedom and the tunable degree of
entanglement that might exist between two quantum subsystems.

\section*{Funding Information}
We acknowledge financial support from the Spanish Ministry of
Economy and Competitiveness through the Severo Ochoa Programme for
Centres of Excellence in R$\&$D (SEV- 2015-0522) and from
Fundaci\'o Privada Cellex. J. P. T. acknowledges support from the
program ICREA Academia (Generalitat de Catalunya). J. M. H.
acknowledges support from the Consejo Nacional de Ciencia y
Tecnolog\'ia  (M\'exico).



\end{document}